# Model of Block Media Taking into Account Internal Friction

**N. I. Aleksandrova**

*Chinakal Institute of Mining of the Siberian Branch of the RAS, Novosibirsk, 630091 Russia*

*e-mail: nialex@misd.ru*

**Abstract** − The block medium is modeled by a discrete-periodic spatial lattice of masses connected by elastic springs and viscous dampers. To describe the viscoelastic behavior of the interblock layers, a rheological model of internal friction with two Maxwell elements and one Voigt element with the quality factor of the material as the determining parameter is proposed. Numerical experiments show that, within the framework of this interlayer model, it is possible to select the viscosity and stiffness of the Maxwell and Voigt elements so that the quality factor of the material differs from the given constant value by no more than 5%. In the one-dimensional case, within the framework of the proposed model, the influence of the quality factor on the dispersion properties of a block medium is studied and it is shown that the greatest effect of the quality factor on the dispersion is observed in the low-frequency part of the spectrum. In the three-dimensional case, within the framework of the proposed model, some geomechanical problems are numerically studied for a block half-space under the action of a surface concentrated vertical load. Namely, the attenuation of the velocity amplitudes of surface blocks was studied depending on the Q-factor under step action and under the action of a Gaussian pulse. In addition, we study a layer on the surface of a half-space under the action of a concentrated vertical impulse load in the case when both the layer and the half-space are block media but have different properties.

Keywords: internal friction, block medium, Lamb problem, half-space, layer on half-space, wave motion, numerical simulation

## 1. INTRODUCTION

According to modern ideas developed in the works of Sadovsky [1] and his followers, rocks are a hierarchical system of blocks of different scale levels. Blocks of the same level are separated by interlayers of rocks with weakened mechanical properties. It was noted in [2, 3] that the sizes of blocks change on a scale from fractions of a rock mass to geoblocks of the earth's crust. In the experimental work [4], it was shown on a two-dimensional model of a block medium (a brick wall), that for a real geomedium it is possible to determine the sizes of the characteristic blocks of the rock mass according to seismic logging data, using the relation discovered in [5] that relates the value of the propagation velocity of a low-frequency wave, the frequency limiting its spectrum, and the longitudinal size of the blocks. As shown in [2, 3, 6], the motion of a block medium can be represented as the motion of rigid blocks due to the deformation of the interlayers. As a result, the dynamics of a block medium can be studied in the pendulum approximation, when it is assumed that the blocks are incompressible, and all deformations and displacements occur due to the compressibility of the interlayers (see, for example, [8, 9]). In [7–9], a block medium is modeled as a three-dimensional lattice of masses connected by Voigt elements in axial and diagonal directions. In [9], the qualitative correspondence of the finite-difference solution of the Lamb problem for a block medium according to this model with the results of field experiments carried out in a limestone quarry is shown. An alternative approach is based on a mathematical model of a block medium with elastic blocks interacting through compliant interlayers [5, 10, 11]. To describe interlayers, various versions of the model were proposed in [10, 11], in which interlayers between elastic blocks can be elastic, viscoelastic, plastic, and porous.

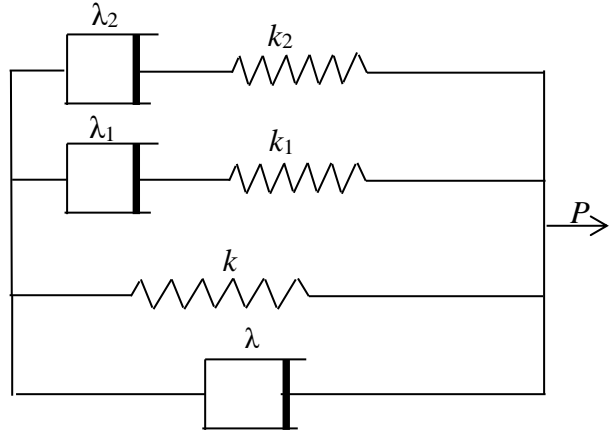

**Fig. 1.** Interlayer model between blocks.

In this article, the 3D model proposed in [8] is modified. In the new model, internal friction in the layers between the blocks is modeled by the Maxwell and Voigt elements. Zener [12] and Biot [13] were among the first to include the Maxwell and Voigt elements in models to describe the inelastic behavior of materials. The model with two Maxwell elements and one Voigt element was first proposed by Biot [13] and also presented by Fang [14]. Many researchers then used various combinations of Maxwell and Voigt elements to account for inelastic losses. In this case, an important task for several decades in each model was to limit the number of relaxation mechanisms to obtain a satisfactory solution, i.e., to obtain an almost constant quality factor of the material $Q$. In [15], a model with internal friction was studied to describe the propagation of waves in homogeneous inelastic media. This model includes two Maxwell elements and one Voigt element. As shown in [15], this model makes it possible to choose the parameters of viscosity and stiffness of these elements, so that the quality factor of the material $Q$ differs from the prescribed constant value by no more than 5% in the frequency range from 3% to 100% of the maximum frequency of interest. An important property of this model is a small number of additional variables with a maximum coverage of the frequency range of interest. The use of a minimum number of Maxwell elements limits the memory and computational resources required in computer programs for the numerical simulation of three-dimensional problems.

Below, to describe the viscoelastic behavior of the interblock layers, we use the internal friction model with two Maxwell elements and one Voigt element with a quality factor $Q$ as the determining parameter. This model is used to solve geomechanical problems of wave propagation.

## 2. ONE-DIMENSIONAL MODEL OF A BLOCK MEDIUM TAKING INTO ACCOUNT INTERNAL FRICTION

Let us first demonstrate this model using the example of a discrete-periodic one-dimensional chain of masses connected by viscoelastic interlayers. The rheological model of the interlayers consists of two Maxwell elements and one Voigt element (Fig. 1). All three elements are connected in parallel. Each Maxwell element consists of a spring and a damper, which are connected in series. The Voigt element also consists of a spring and a damper, but they are connected in parallel. Here $k$, $k_1$, $k_2$ are the spring stiffnesses in the Voigt element and in two Maxwell elements, respectively, $\lambda$, $\lambda_1$, $\lambda_2$ are the damper viscosities in the Voigt element and in two Maxwell elements.

The equations of one-dimensional motion of blocks with this rheological model of interlayers have the following form:

$$M\ddot{u}_j = K[(u_{j+1} - 2u_j + u_{j-1}) + \beta(\dot{u}_{j+1} - 2\dot{u}_j + \dot{u}_{j-1}) \quad (1.1)$$

$$-\alpha_1\gamma_1(\psi_{j+1} - 2\psi_j + \psi_{j-1}) - \alpha_2\gamma_2(\varphi_{j+1} - 2\varphi_j + \varphi_{j-1})], \quad j = 1, 2, ...,$$

$$M\ddot{u}_0 = K[(u_1 - u_0) + \beta(\dot{u}_1 - \dot{u}_0) - \alpha_1\gamma_1(\psi_1 - \psi_0) - \alpha_2\gamma_2(\varphi_1 - \varphi_0)] + P(t),$$

$$\psi_j = e^{-\gamma_1 t}\int_0^t e^{\gamma_1 \tau} u_j(\tau)d\tau, \quad \varphi_j = e^{-\gamma_2 t}\int_0^t e^{\gamma_2 \tau} u_j(\tau)d\tau,$$

$$\alpha_1 = \frac{k_1}{K}, \quad \alpha_2 = \frac{k_2}{K}, \quad \gamma_1 = \frac{k_1}{\lambda_1}, \quad \gamma_2 = \frac{k_2}{\lambda_2}, \quad \beta = \frac{\lambda}{K}, \quad K = k + k_1 + k_2.$$

Here, $u_j$ are the displacements of rigid blocks, $P(t)$ is the actual load applied to the block $j = 0$; $K$ is the total stiffness of the springs. In (1.1), two additional variables $\varphi_j$ and $\psi_j$ are introduced, which depend on the displacement function $u_j$.

Let us apply the Fourier transform in time to the equations of motion (1.1). In the frequency domain, the force acting on the block from the side of the interlayer can be expressed by the formula:

$$\tilde{P}(\omega) = \tilde{F}(\omega)\tilde{u}(\omega),$$

where

$$\tilde{F}(\omega) = K\left[\left(1 - \frac{\alpha_1\gamma_1^2}{\omega^2 + \gamma_1^2} - \frac{\alpha_2\gamma_2^2}{\omega^2 + \gamma_2^2}\right) + i\omega\left(\beta + \frac{\alpha_1\gamma_1}{\omega^2 + \gamma_1^2} + \frac{\alpha_2\gamma_2}{\omega^2 + \gamma_2^2}\right)\right]$$

is a normalized impedance function [16].

Internal attenuation in a medium is usually determined by the quality factor $Q(\omega)$ or its reciprocal $Q^{-1}(\omega)$, which is described by the formula:

$$Q^{-1}(\omega) = \frac{\text{Im}\,\tilde{F}(\omega)}{\text{Re}\,\tilde{F}(\omega)} = \frac{\omega\left[\beta + \alpha_1\gamma_1/(\omega^2 + \gamma_1^2) + \alpha_2\gamma_2/(\omega^2 + \gamma_2^2)\right]}{1 - \alpha_1\gamma_1^2/(\omega^2 + \gamma_1^2) - \alpha_2\gamma_2^2/(\omega^2 + \gamma_2^2)}. \quad (1.2)$$

For soils and rock materials, it is customary to assume that the target quality factor of the material remains constant over a wide range of frequencies of interest, i.e., $Q_0(\omega) = Q_0$, see [17].

To relate the parameters $\alpha_1$, $\alpha_2$, $\gamma_1$, $\gamma_2$ and $\beta$ of the rheological model to $Q^{-1}(\omega)$ using simple frequency independent approximations, we introduce new dimensionless parameters $\hat{\alpha}_1$, $\hat{\alpha}_2$, $\hat{\gamma}_1$, $\hat{\gamma}_2$, $\hat{\beta}$ and the frequency $\hat{\omega}$ using the formulas:

$$\hat{\alpha}_1 = \alpha_1 Q_0, \quad \hat{\alpha}_2 = \alpha_2 Q_0, \quad \hat{\gamma}_1 = \gamma_1/\omega_{\max}, \quad \hat{\gamma}_2 = \gamma_2/\omega_{\max}, \quad \hat{\beta} = \beta\omega_{\max}Q_0, \quad \hat{\omega} = \omega/\omega_{\max}, \quad (1.3)$$

where $\omega_{\max}$ is the maximum angular frequency that is of interest for modeling.

In expression (1.2) for the coefficient $Q^{-1}(\omega)$, we pass to the normalized parameters (1.3). Then formula (1.2) takes the form:

$$\frac{Q^{-1}(\hat{\omega}, Q_0^{-1})}{Q_0^{-1}} = \frac{Q_0\hat{\omega}\left[\hat{\beta} + \hat{\alpha}_1\hat{\gamma}_1/(\hat{\omega}^2 + \hat{\gamma}_1^2) + \hat{\alpha}_2\hat{\gamma}_2/(\hat{\omega}^2 + \hat{\gamma}_2^2)\right]}{1 - \hat{\alpha}_1\hat{\gamma}_1^2/[Q_0(\hat{\omega}^2 + \hat{\gamma}_1^2)] - \hat{\alpha}_2\hat{\gamma}_2^2/[Q_0(\hat{\omega}^2 + \hat{\gamma}_2^2)]}.$$

**Table 1.** Parameter values corresponding to different values of $Q_0^{-1}$.

| $Q_0^{-1}$ | $\hat{\alpha}_1$ | $\hat{\alpha}_2$ | $\hat{\gamma}_1$ | $\hat{\gamma}_2$ | $\hat{\beta}$ |
|---|---|---|---|---|---|
| 0.2 | 0.28 | 0.28 | 0.026 | 0.23 | 0.14 |
| 0.1 | 0.14 | 0.14 | 0.03 | 0.23 | 0.07 |
| 0.05 | 0.072 | 0.072 | 0.027 | 0.22 | 0.035 |
| 0.02 | 0.029 | 0.029 | 0.026 | 0.22 | 0.014 |
| 0.01 | 0.014 | 0.014 | 0.031 | 0.222 | 0.0071 |

From this formula, we can see that the coefficients $\hat{\alpha}_1$, $\hat{\alpha}_2$, $\hat{\gamma}_1$, $\hat{\gamma}_2$, $\hat{\beta}$ do not depend on $\omega_{max}$. It follows that for a given constant $Q_0^{-1}$, they should only be calculated once. At large values of the target quality factor, i.e. $Q_0^{-1} \to 0$, the parameters become essentially independent of $Q_0$.

The problem of determining the parameters $\hat{\alpha}_1$, $\hat{\alpha}_2$, $\hat{\gamma}_1$, $\hat{\gamma}_2$, $\hat{\beta}$ is solved with a tolerance, so that the actual quality factor remains close enough to the target value. This tolerance, which is taken as 5% of the target value, is carried out between 4% and 100% of $\omega_{max}$. That is, for a given quality factor $Q_0$, parameters $\hat{\alpha}_1$, $\hat{\alpha}_2$, $\hat{\gamma}_1$, $\hat{\gamma}_2$, are selected by direct checking on a computer so that the condition

$$\left[ Q^{-1}(\hat{\omega}, Q_0^{-1}) Q_0 - 1 \right]^2 < 0.05^2$$

is satisfied for each $\hat{\omega}$, that satisfies the inequality $0.04 \leq \hat{\omega} \leq 1$. The found values of the parameters corresponding to five values of the target quality factor are presented in Table 1.

For these parameter values, the actual Q-factors of the material, normalized to $Q_0^{-1}$, are shown as a function of frequency in Fig. 2. As can be seen in Fig. 2, it is possible to choose the parameters $\hat{\alpha}_1$, $\hat{\alpha}_2$, $\hat{\gamma}_1$, $\hat{\gamma}_2$, $\hat{\beta}$ so that the tolerance, which is taken as 5% of the target value, is performed within the range from 4% to 100% of $\omega_{max}$.

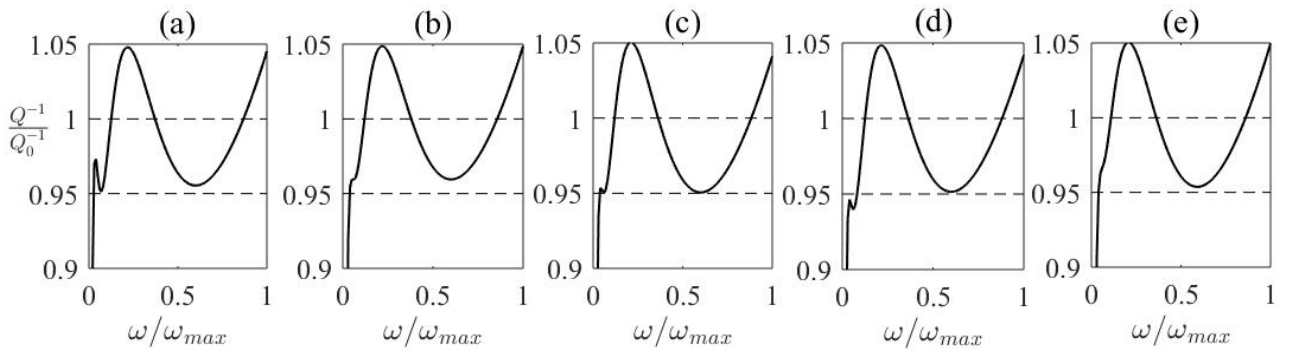

**Fig. 2.** Dependences of the quality factor of the material on the frequency: (a) $Q_0^{-1} = 0.2$; (b) $Q_0^{-1} = 0.1$; (c) $Q_0^{-1} = 0.05$; (d) $Q_0^{-1} = 0.02$; (e) $Q_0^{-1} = 0.01$.

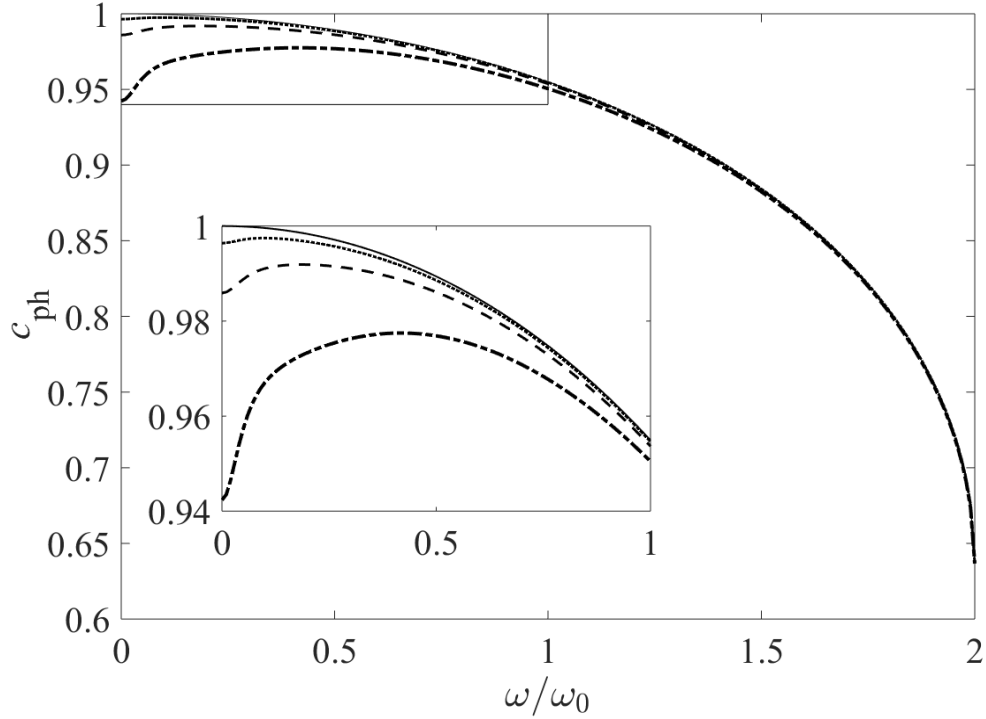

**Fig. 3.** Frequency dependence of the phase velocity for a one-dimensional model of a block medium. Solid curves correspond to $Q_0^{-1}=0$, dotted curves correspond to $Q_0^{-1}=0.05$, dashed curves correspond to $Q_0^{-1}=0.1$, dash-dotted curves correspond to $Q_0^{-1}=0.2$.

Another important aspect of internal friction models is the dispersion observed in the time response of the system. For the discrete equation (1.1), the dispersion dependence of the wave number $q$ on the frequency $\omega$ has the form:

$$\sin^2\left(\frac{ql}{2}\right) = \frac{M\omega^2}{4\tilde{F}(\omega)}. \tag{1.4}$$

Here, $l$ is the length of the springs, $q$ is the wavenumber. Using formula (1.4) and the relation $c_{ph}(\omega)=\omega/q$, where $c_{ph}$ is the phase velocity, we can determine the frequency dependence of the phase velocity for the one-dimensional model of a block medium (1.1). If we put $\omega\to 0$, then from (1.4) we get: $c_{ph}(0)=l\sqrt{K(1-\alpha_1-\alpha_2)/M}$. As follows from this formula and Table 1, the phase velocity of infinitely long waves decreases with increasing $Q_0^{-1}$.

Figure 3 shows the dependencies $c_{ph}(\omega)$ for different values of the quality factor of the material $Q_0^{-1}$. Here and below we assume: $l=1$, $M=1$, $K=1$. The rectangle shows the same curves on an enlarged scale in the low-frequency zone. Note that in the discrete case we have: $\omega_{max}=2\omega_0$, where $\omega_0=\sqrt{K/M}$. It can be seen that the greatest influence of the quality factor on the dispersion is observed in the low-frequency part of the spectrum. In addition, it can be seen that as increases $Q_0^{-1}$, the maximum perturbation propagation velocity decreases. In the high-frequency part of the spectrum, the phase velocity is practically independent of the quality factor.

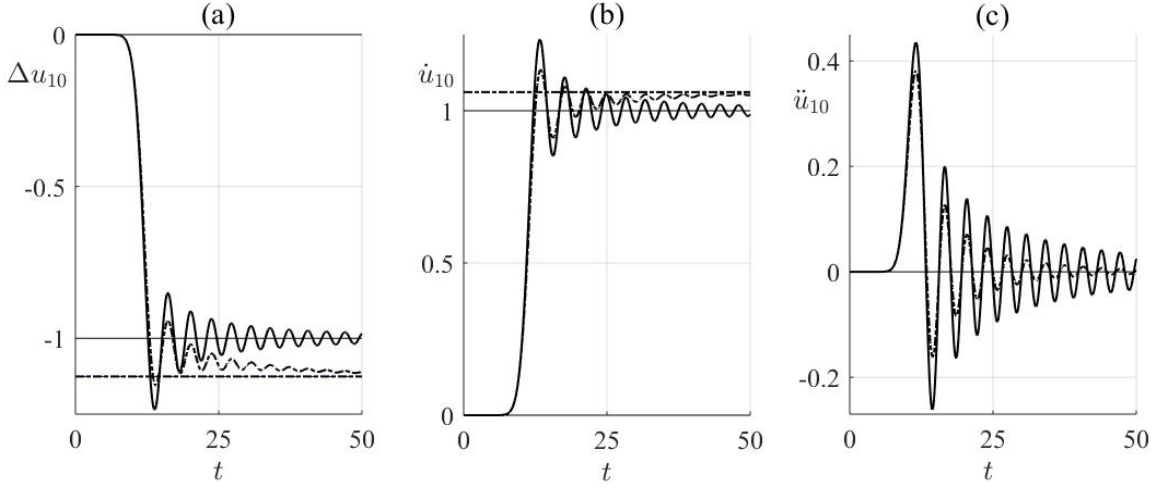

**Fig. 4.** Dependences on the time of deformation of the interlayer (a), velocity (b) and acceleration (c) of the 10th block. Solid curves correspond to $Q_0^{-1} = 0$, dash-dotted curves correspond to $Q_0^{-1} = 0.2$.

Figure 4 shows the results of numerical calculations of a nonstationary wave process in a chain of masses under the action of a load $P(t) = H(t)$, where $H(t)$ is the Heaviside step function. The calculations were carried out using a one-dimensional model with internal friction (1.1) for a block with the number $j = 10$. The solid curves correspond to $Q_0^{-1} = 0$, the dash-dotted curves correspond to $Q_0^{-1} = 0.2$. Figure 4a shows the deformations of the interlayer $\Delta u_j = u_{j+1} - u_j$, Fig. 4b shows the velocities and Fig. 4c shows the accelerations as a function of time.

Horizontal lines correspond to static values: $\Delta u_{st} = -1/(1 - \alpha_1 - \alpha_2)$ in Fig. 4a, $\dot{u}_{st} = 1/\sqrt{1 - \alpha_1 - \alpha_2}$ in Fig. 4b. The amplitudes of the deformation of the interlayers and the velocities of the blocks tend to these values at large values of time. It can be seen that with increasing $Q_0^{-1}$ the attenuation of the maximum amplitudes of interlayer deformations, velocities, and accelerations of blocks at the low-frequency wave front increases, and the amplitude of high-frequency oscillations $\Delta u_j$ and $\dot{u}_j$ decreases relative to their static values behind the wave front.

## 3. THREE-DIMENSIONAL MODEL OF A BLOCK MEDIUM TAKING INTO ACCOUNT INTERNAL FRICTION

The block medium is modeled by a homogeneous three-dimensional lattice consisting of point masses connected by springs and dampers in the directions of the *x*, *y*, *z* axes and in the diagonal directions of the *xy*, *xz*, *yz* planes, as shown in Fig. 5a. The notation used here is: *u*, *v* are horizontal displacements in the *x*, *y* directions; *w* are vertical displacements in the *z* direction; *n*, *m*, *k* are block numbers in *x*, *y*, *z* directions. A vertical concentrated load *P* is applied on the surface of the block half-space at the origin of coordinates (Fig. 5b).

As a rheological model of interlayers, the interlayer model with two Maxwell elements and one Voigt element, which has just been demonstrated on the example of a one-dimensional block medium, is used. Further, we will assume that the stiffness of the springs and the viscosity of the dampers in the axial and diagonal directions are the same. In addition, we will assume that the parameters *M*, *k*, $k_1$, $k_2$, $\lambda$, $\lambda_1$, $\lambda_2$ have the same values at all points of the block medium.

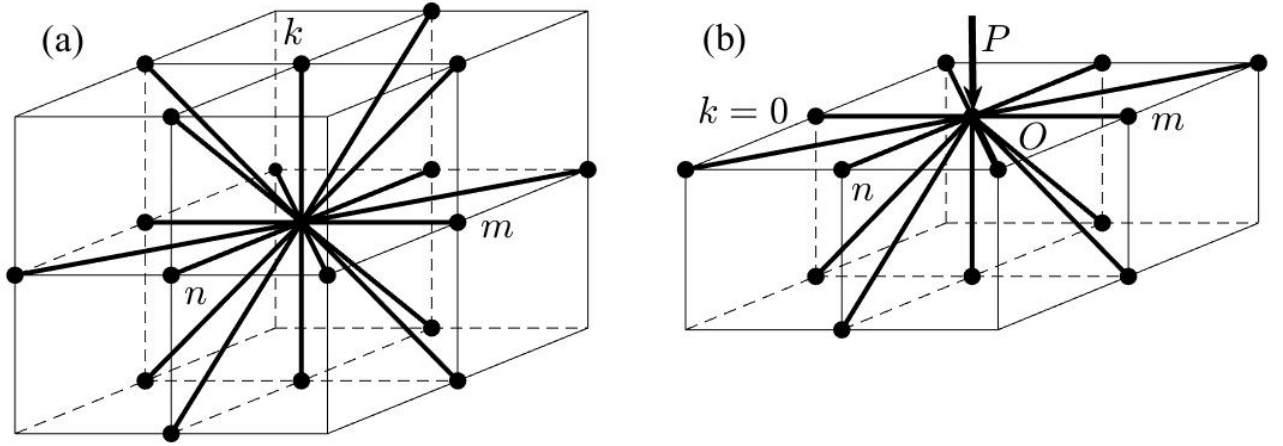

**Fig. 5.** Scheme of connecting the masses by springs and dampers (a) inside the half-space and (b) on the surface of the half-space.

Taking into account internal friction and notation

$$\Lambda_{nn} f_{n,m,k} = f_{n+1,m,k} - 2 f_{n,m,k} + f_{n-1,m,k},$$

$$\Phi_{nm} f_{n,m,k} = (f_{n+1,m+1,k} + f_{n-1,m-1,k} - 4 f_{n,m,k} + f_{n+1,m-1,k} + f_{n-1,m+1,k})/2,$$

$$\Psi_{nm} f_{n,m,k} = (f_{n+1,m+1,k} + f_{n-1,m-1,k} - f_{n+1,m-1,k} - f_{n-1,m+1,k})/2,$$

$$\psi^u_{n,m,k} = e^{-\gamma_1 t}\int_0^t e^{\gamma_1 \tau} u_{n,m,k}(\tau)d\tau, \ \psi^v_{n,m,k} = e^{-\gamma_1 t}\int_0^t e^{\gamma_1 \tau} v_{n,m,k}(\tau)d\tau, \ \psi^w_{n,m,k} = e^{-\gamma_1 t}\int_0^t e^{\gamma_1 \tau} w_{n,m,k}(\tau)d\tau,$$

$$\varphi^u_{n,m,k} = e^{-\gamma_2 t}\int_0^t e^{\gamma_2 \tau} u_{n,m,k}(\tau)d\tau, \ \varphi^v_{n,m,k} = e^{-\gamma_2 t}\int_0^t e^{\gamma_2 \tau} v_{n,m,k}(\tau)d\tau, \ \varphi^w_{n,m,k} = e^{-\gamma_2 t}\int_0^t e^{\gamma_2 \tau} w_{n,m,k}(\tau)d\tau$$

the equations of motion of a block with coordinates $n$, $m$, $k$, located inside the half-space ($k < 0$), have the form:

$$\begin{aligned}
M\ddot{u}_{n,m,k} = & K\{(\Lambda_{nn} + \Phi_{nk} + \Phi_{nm})u_{n,m,k} + \Psi_{nm} v_{n,m,k} + \Psi_{nk} w_{n,m,k} \\
& + \beta[(\Lambda_{nn} + \Phi_{nk} + \Phi_{nm})\dot{u}_{n,m,k} + \Psi_{nm}\dot{v}_{n,m,k} + \Psi_{nk}\dot{w}_{n,m,k}] \\
& - \alpha_1\gamma_1[(\Lambda_{nn} + \Phi_{nk} + \Phi_{nm})\psi^u_{n,m,k} + \Psi_{nm}\psi^v_{n,m,k} + \Psi_{nk}\psi^w_{n,m,k}] \\
& - \alpha_2\gamma_2[(\Lambda_{nn} + \Phi_{nk} + \Phi_{nm})\varphi^u_{n,m,k} + \Psi_{nm}\varphi^v_{n,m,k} + \Psi_{nk}\varphi^w_{n,m,k}]\},
\end{aligned} \quad (3.1)$$

$$\begin{aligned}
M\ddot{v}_{n,m,k} = & K\{\Psi_{nm} u_{n,m,k} + (\Lambda_{mm} + \Phi_{mk} + \Phi_{nm})v_{n,m,k} + \Psi_{mk} w_{n,m,k} \\
& + \beta[\Psi_{nm}\dot{u}_{n,m,k} + (\Lambda_{mm} + \Phi_{mk} + \Phi_{nm})\dot{v}_{n,m,k} + \Psi_{mk}\dot{w}_{n,m,k}] \\
& - \alpha_1\gamma_1[\Psi_{nm}\psi^u_{n,m,k} + (\Lambda_{mm} + \Phi_{mk} + \Phi_{nm})\psi^v_{n,m,k} + \Psi_{mk}\psi^w_{n,m,k}] \\
& - \alpha_2\gamma_2[\Psi_{nm}\varphi^u_{n,m,k} + (\Lambda_{mm} + \Phi_{mk} + \Phi_{nm})\varphi^v_{n,m,k} + \Psi_{mk}\varphi^w_{n,m,k}]\},
\end{aligned} \quad (3.2)$$

$$\begin{aligned}
M\ddot{w}_{n,m,k} = & K\{\Psi_{nk} u_{n,m,k} + \Psi_{mk} v_{n,m,k} + (\Lambda_{kk} + \Phi_{mk} + \Phi_{nk})w_{n,m,k} \\
& + \beta[\Psi_{nk}\dot{u}_{n,m,k} + \Psi_{mk}\dot{v}_{n,m,k} + (\Lambda_{kk} + \Phi_{mk} + \Phi_{nk})\dot{w}_{n,m,k}] \\
& - \alpha_1\gamma_1[\Psi_{nk}\psi^u_{n,m,k} + \Psi_{mk}\psi^v_{n,m,k} + (\Lambda_{kk} + \Phi_{mk} + \Phi_{nk})\psi^w_{n,m,k}] \\
& - \alpha_2\gamma_2[\Psi_{nk}\varphi^u_{n,m,k} + \Psi_{mk}\varphi^v_{n,m,k} + (\Lambda_{kk} + \Phi_{mk} + \Phi_{nk})\varphi^w_{n,m,k}]\}.
\end{aligned} \quad (3.3)$$

Taking into account the notations

$$\Lambda_k^- f_{n,m,0} = f_{n+1,m,-1} - f_{n,m,0}, \quad \Phi_{nk}^- f_{n,m,0} = (f_{n-1,m,-1} - 2f_{n,m,0} + f_{n+1,m,-1})/2,$$

$$\Phi_{mk}^- f_{n,m,0} = (f_{n,m-1,-1} - 2f_{n,m,0} + f_{n,m+1,-1})/2, \quad \Psi_{nk}^- f_{n,m,0} = (f_{n-1,m,-1} - f_{n+1,m,-1})/2,$$

$$\Psi_{mk}^- f_{n,m,0} = (f_{n,m-1,-1} - f_{n,m+1,-1})/2$$

the equations of motion of a block with coordinates $n$, $m$, 0, located on the free surface of a half-space, have the form:

$$M\ddot{u}_{n,m,0} = K\{(\Lambda_{nn} + \Phi_{nk}^- + \Phi_{nm})u_{n,m,0} + \Psi_{nm}v_{n,m,0} + \Psi_{nk}^- w_{n,m,0}$$
$$+ \beta[(\Lambda_{nn} + \Phi_{nk}^- + \Phi_{nm})\dot{u}_{n,m,0} + \Psi_{nm}\dot{v}_{n,m,0} + \Psi_{nk}^- \dot{w}_{n,m,0}]$$
$$- \alpha_1\gamma_1[(\Lambda_{nn} + \Phi_{nk}^- + \Phi_{nm})\psi^u_{n,m,0} + \Psi_{nm}\psi^v_{n,m,0} + \Psi_{nk}^- \psi^w_{n,m,0}]$$
$$- \alpha_2\gamma_2[(\Lambda_{nn} + \Phi_{nk}^- + \Phi_{nm})\varphi^u_{n,m,0} + \Psi_{nm}\varphi^v_{n,m,0} + \Psi_{nk}^- \varphi^w_{n,m,0}]\}, \quad (3.4)$$

$$M\ddot{v}_{n,m,0} = K\{\Psi_{nm}u_{n,m,0} + (\Lambda_{mm} + \Phi_{mk}^- + \Phi_{nm})v_{n,m,0} + \Psi_{mk}^- w_{n,m,0}$$
$$+ \beta[\Psi_{nm}\dot{u}_{n,m,0} + (\Lambda_{mm} + \Phi_{mk}^- + \Phi_{nm})\dot{v}_{n,m,0} + \Psi_{mk}^- \dot{w}_{n,m,0}]$$
$$- \alpha_1\gamma_1[\Psi_{nm}\psi^u_{n,m,0} + (\Lambda_{mm} + \Phi_{mk}^- + \Phi_{nm})\psi^v_{n,m,0} + \Psi_{mk}^- \psi^w_{n,m,0}]$$
$$- \alpha_2\gamma_2[\Psi_{nm}\varphi^u_{n,m,0} + (\Lambda_{mm} + \Phi_{mk}^- + \Phi_{nm})\varphi^v_{n,m,0} + \Psi_{mk}^- \varphi^w_{n,m,0}]\}, \quad (3.5)$$

$$M\ddot{w}_{n,m,0} = K\{\Lambda_k^- w_{n,m,0} + \Psi_{nk}^- u_{n,m,0} + \Psi_{mk}^- v_{n,m,0} + (\Phi_{mk}^- + \Phi_{nk}^-)w_{n,m,0}$$
$$+ \beta[\Lambda_k^- \dot{w}_{n,m,0} + \Psi_{nk}^- \dot{u}_{n,m,0} + \Psi_{mk}^- \dot{v}_{n,m,0} + (\Phi_{mk}^- + \Phi_{nk}^-)\dot{w}_{n,m,0}]$$
$$- \alpha_1\gamma_1[\Lambda_k^- \psi^w_{n,m,0} + \Psi_{nk}^- \psi^u_{n,m,0} + \Psi_{mk}^- \psi^v_{n,m,0} + (\Phi_{mk}^- + \Phi_{nk}^-)\psi^w_{n,m,0}]$$
$$- \alpha_2\gamma_2[\Lambda_k^- \varphi^w_{n,m,0} + \Psi_{nk}^- \varphi^u_{n,m,0} + \Psi_{mk}^- \varphi^v_{n,m,0} + (\Phi_{mk}^- + \Phi_{nk}^-)\varphi^w_{n,m,0}]\} + P(t)\delta_{n0}\delta_{m0}. \quad (3.6)$$

Here, $\delta_{n0}$ is the Kronecker symbol, $P(t)$ is the amplitude of the acting surface load. The initial conditions are zero.

As shown in [8], Eqs. (3.1) – (3.6), which describe the motion of a block half-space with elastic layers ($\alpha_1 = 0$, $\alpha_2 = 0$, $\gamma_1 = 0$, $\gamma_2 = 0$, $\beta = 0$), correspond to the velocities of longitudinal ($c_p$), shear ($c_s$) and Rayleigh ($c_R$) infinitely long waves:

$$c_p = l\sqrt{\frac{3K}{M}}, \quad c_s = l\sqrt{\frac{K}{M}}, \quad c_R = l\sqrt{\frac{2K}{M}\left(1 - \frac{1}{\sqrt{3}}\right)}. \quad (3.7)$$

Using the block medium model (3.1)–(3.6), two problems of wave propagation in a three-dimensional block medium are numerically solved taking into account internal friction with a concentrated vertical action applied at the origin. In Problem 1, the load acts on the surface of the block half-space. In Problem 2, the load acts on the surface of a block layer lying on a block half-space, and the properties of the layer and half-space are different.

For these problems, the response of a block medium to two types of acting load is studied: (a) step action, i.e. $P(t) = P_0 H(t)$; (b) Gaussian momentum, i.e. $P(t) = P_0 \exp\left[-(t - 4\sigma)^2/(2\sigma^2)\right]$.

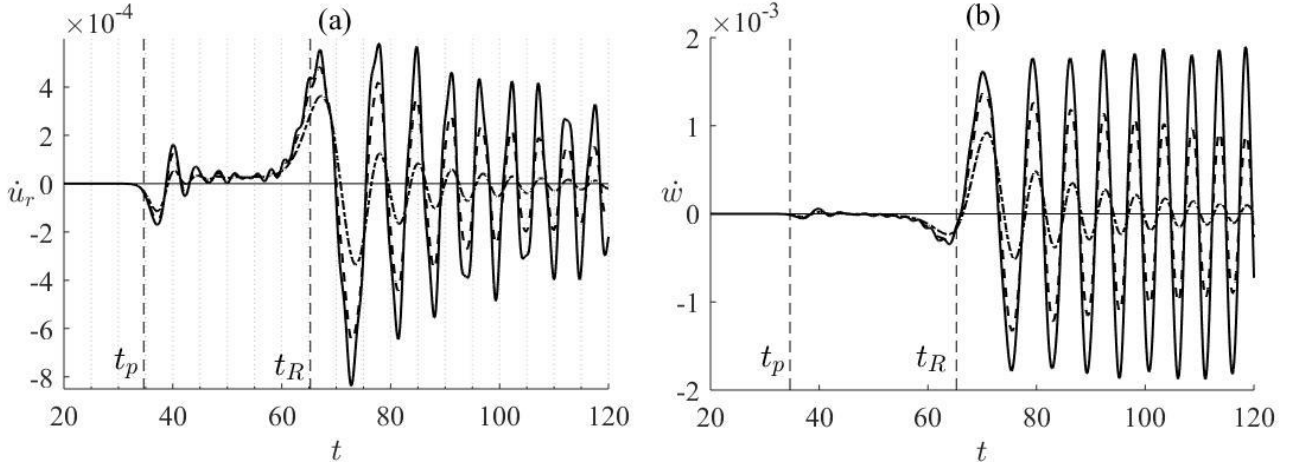

**Fig. 6.** Time dependences of the radial (a) and vertical (b) velocities of the blocks at the point (60, 0, 0) under the action of a step load. Solid curves correspond to $Q_0^{-1} = 0$, dashed curves correspond to $Q_0^{-1} = 0.1$, dash-dotted curves correspond to $Q_0^{-1} = 0.2$.

## 4. DIFFERENCE SCHEME

Equations (3.1)–(3.6) with zero initial data were solved by the finite difference method using an explicit scheme. For the second time derivatives, the central-difference approximation of the second order of accuracy was used, for the first time derivatives, the "backward" difference of the first order of accuracy was used:

$$\ddot{u}_{n,m,k} \approx (u_{n,m,k}^{s+1} - 2u_{n,m,k}^{s} + u_{n,m,k}^{s-1})/\tau^2, \qquad s = 0,1,2,...,$$

$$\dot{u}_{n,m,k} \approx (u_{n,m,k}^{s} - u_{n,m,k}^{s-1})/\tau, \qquad s = 0,1,2,....$$

Here, $\tau$ is the step of the difference grid in time, $u_{n,m,k}^{s}$ is the displacement value $u_{n,m,k}(t)$ at moment of time $t = s\tau$, $s$ is the layer number in time in the finite difference scheme. For additional functions, the following approximations were used:

$$\psi_{n,m,k}^{u,s} = \tau[u_{n,m,k}^{s}(1-\gamma_1\tau) + u_{n,m,k}^{s-1}]/2 + \psi_{n,m,k}^{u,s-1}e^{-\gamma_1\tau},$$

$$\varphi_{n,m,k}^{u,s} = \tau[u_{n,m,k}^{s}(1-\gamma_2\tau) + u_{n,m,k}^{s-1}]/2 + \varphi_{n,m,k}^{u,s-1}e^{-\gamma_2\tau}.$$

## 5. RESULTS OF NUMERICAL CALCULATIONS OF PROBLEM 1 TAKING INTO ACCOUNT INTERNAL FRICTION

This section presents the results of numerical calculations of perturbations in the Lamb problem with allowance for internal friction, carried out for $P_0 = 1$ and $\tau = \pi/20$. The mass of the blocks, the length of the springs, and the total stiffness of the layers are taken as units: $M = 1$, $l = 1$, $K = 1$.

Figures 6–9 show the results of calculations of the vertical $\dot{w} = \dot{w}_{n,m,k}$ and radial $\dot{u}_r = (n\dot{u}_{n,m,k} + m\dot{v}_{n,m,k})/(n^2 + m^2)^{1/2}$ velocities of the blocks in the cylindrical coordinate system on the surface of the block half-space ($k = 0$) on the $m = 0$ axis.

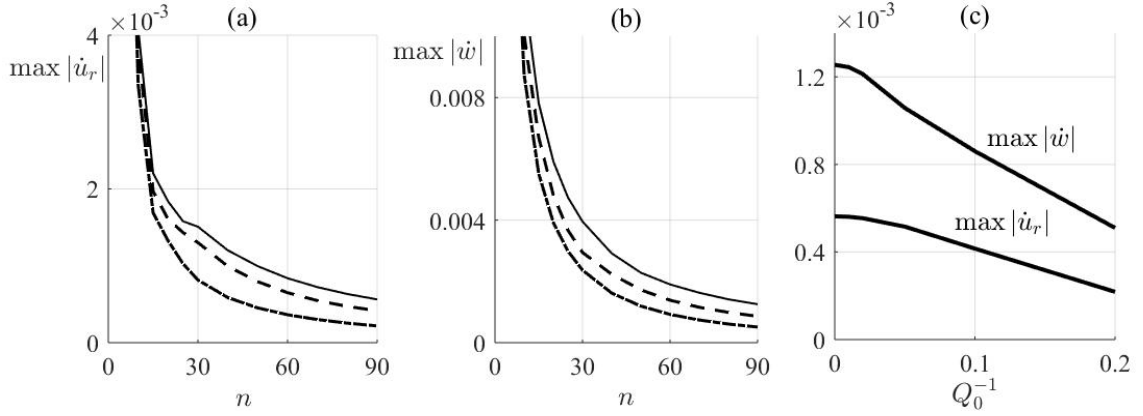

**Fig. 7.** Dependences on the coordinate $n$ of the maximum values of the absolute values of the amplitudes of the radial (a) and vertical (b) velocities of the blocks on the surface of the block half-space; (c) Dependences on the Q-factor $Q_0^{-1}$ of the maximum values of the absolute values of the radial and vertical velocities of the blocks at the point (90, 0, 0). Solid curves correspond to $Q_0^{-1}=0$, dashed curves correspond to $Q_0^{-1}=0.1$, dash-dotted curves correspond to $Q_0^{-1}=0.2$.

Figure 6 shows the graphs of the time dependence of the radial and vertical velocities of blocks at the point (60, 0, 0), calculated under the action of a step load at various values of the quality factor of the material $Q_0^{-1}$. The vertical lines correspond to the arrival times of longitudinal waves $t_p = n/c_p$ and Rayleigh waves $t_R = n/c_R$, where $c_p$ and $c_R$ are determined by formulas (3.7). It can be seen that with increasing $Q_0^{-1}$, the attenuation of the amplitude of block velocities increases for the entire frequency spectrum.

Figure 7 shows the results of calculations of the maximum values of the absolute values of the amplitudes of the radial and vertical velocities of the blocks under the action of a step load in time. Figures 7a and 7b show the dependences of these quantities on the coordinate $n$ for various values of the quality factor $Q_0^{-1}$. Figure 7c shows the dependence on the quality factor $Q_0^{-1}$ at the point (90, 0, 0). It can be seen that as the coefficient $Q_0^{-1}$ increases, the maximum values of the absolute values of the amplitudes $\dot{u}_r$ and $\dot{w}$ fall faster with increasing $n$ and their dependence on the coefficient $Q_0^{-1}$ is close to linear.

Figure 8 shows the graphs of the time dependence of the radial and vertical velocities of the blocks at the point (60, 0, 0), calculated for different values of the quality factor $Q_0^{-1}$ under the action of a Gaussian pulse ($\sigma=5$). The vertical lines correspond to the moments of arrival of longitudinal and Rayleigh waves. It can be seen that with increasing $Q_0^{-1}$ the amplitudes of the block velocities decrease and the time for the onset of the maximum amplitudes of the velocities increases. The latter is explained by a decrease in the speed of low-frequency waves, which can be seen in Fig. 2.

Figure 9 shows the results of calculations of the maximum values of the absolute values of the amplitudes of the radial and vertical velocities of blocks on the surface of a block half-space under the action of a Gaussian pulse ($\sigma=5$). Figures 9a and 9b shows the dependences of these quantities on the coordinate $n$ for different on values of the quality factor $Q_0^{-1}$, Fig. 9c shows the dependence the quality factor at the point (100, 0, 0).

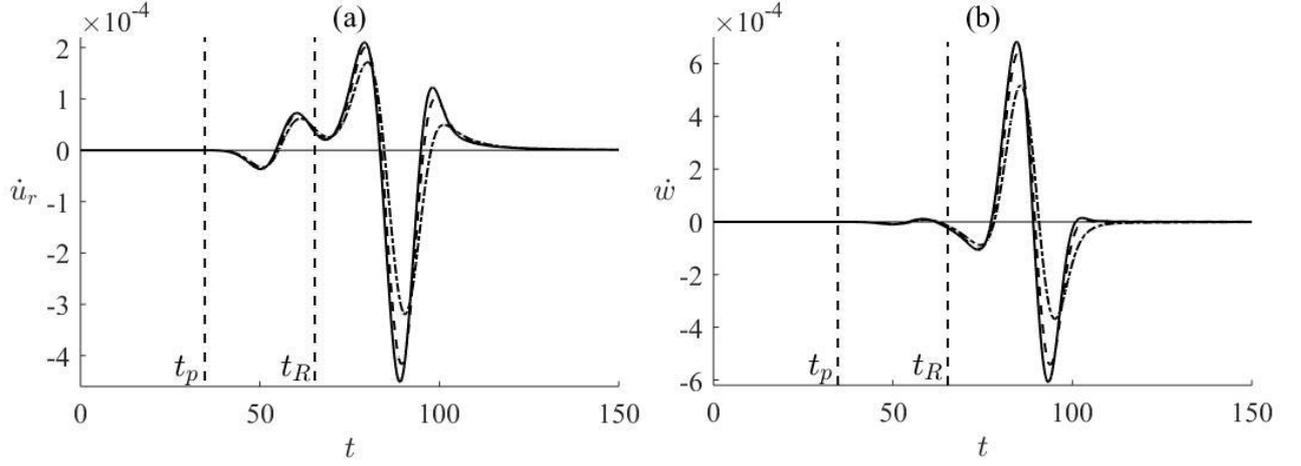

**Fig. 8.** Time dependences of the radial (a) and vertical (b) velocities of the blocks at the point (60, 0, 0) under the action of a Gaussian pulse. Solid curves correspond to $Q_0^{-1}=0$, dashed curves correspond to $Q_0^{-1}=0.1$, dash-dotted curves correspond to $Q_0^{-1}=0.2$.

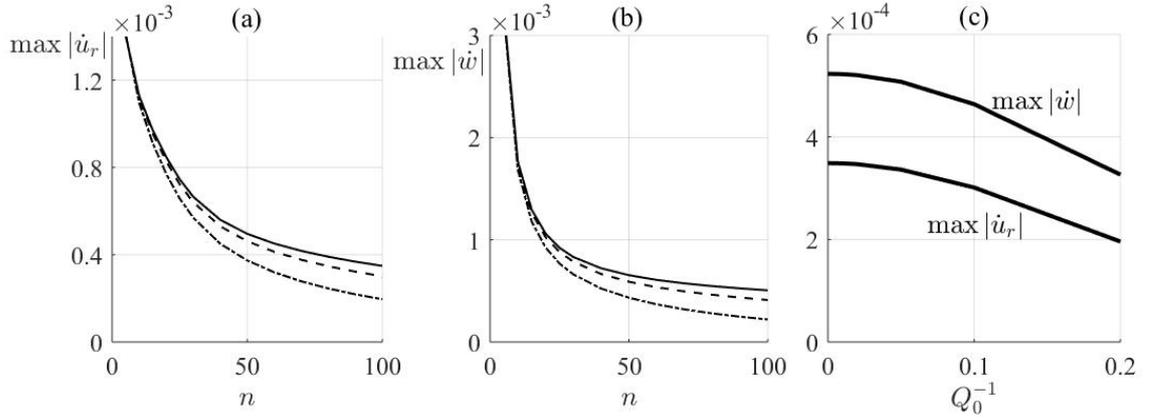

**Fig. 9.** Dependences on the coordinate $n$ of the maximum values of the absolute values of the amplitudes of the radial (a) and vertical (b) velocities of the blocks at different values of the quality factor $Q_0^{-1}$; (c) dependences on the Q-factor of the maximum values of the absolute values of the amplitudes of the radial and vertical velocities at the point (100, 0, 0). Solid curves correspond to $Q_0^{-1}=0$, dashed curves correspond to $Q_0^{-1}=0.1$, dash-dotted curves correspond to $Q_0^{-1}=0.2$.

For the graphs presented in Figs. 9a and 9b the following approximations by the power function of the results of numerical calculations were obtained:

$Q_0^{-1}=0.2:\ \max|\dot{u}_r|=0.0055n^{-0.695},\ \max|\dot{w}|=0.0102n^{-0.738}$,

$Q_0^{-1}=0.1:\ \max|\dot{u}_r|=0.0039n^{-0.546},\ \max|\dot{w}|=0.0074n^{-0.607}$,

$Q_0^{-1}=0:\ \max|\dot{u}_r|=0.0035n^{-0.495},\ \max|\dot{w}|=0.0067n^{-0.562}$.

Analysis of these approximate functions and graphs in Figs. 9a and 9b shows that with increasing coefficient $Q_0^{-1}$, the damping rate of the maximum modulo values $\dot{u}_r$, $\dot{w}$ increases with distance from the impact site. Figure 9c shows that the dependence of the maximum amplitudes of block velocities on the coefficient $Q_0^{-1}$ is close to parabolic.

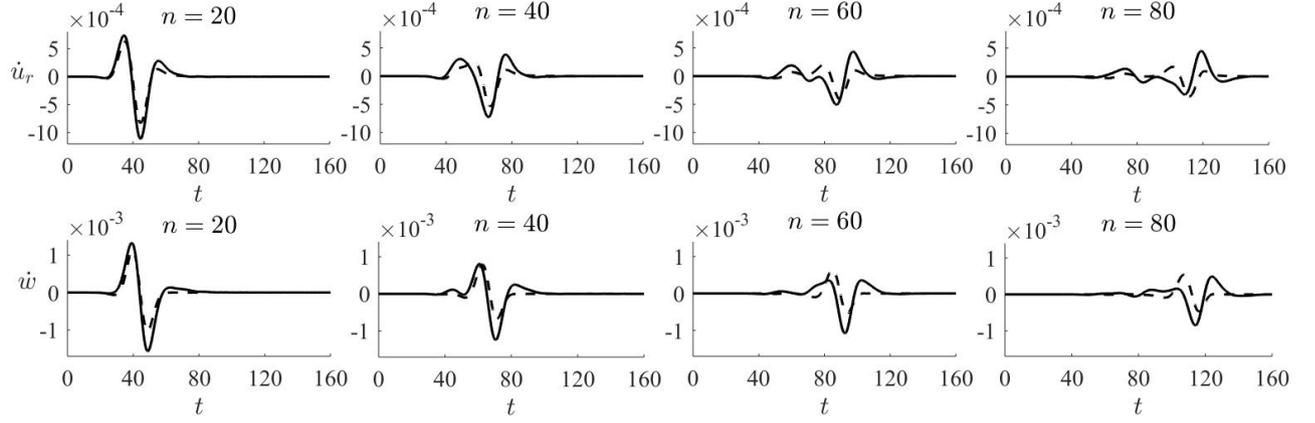

**Fig. 10.** Time dependences of the radial and vertical velocities of blocks on the surface of a block half-space. The dashed curves are the block half-space, the solid curves are the block layer on the block half-space. The top row of graphs is the radial velocities of the blocks, the bottom row is the vertical velocities.

## 6. RESULTS OF NUMERICAL CALCULATIONS OF PROBLEMS 1 AND 2 TAKING INTO ACCOUNT INTERNAL FRICTION

Figure 10 below shows the results of numerical calculations of two problems, taking into account internal friction under the action of a vertical concentrated Gaussian pulse, firstly, on the surface of a block half-space (Problem 1, shown by dashed curves) and, secondly, on the surface of a block layer lying on a block half-space with different properties (Problem 2, shown by solid curves).

Figure 10 shows the time dependences of the radial and vertical velocities of the blocks on the surface of the block half-space located at different distances from the impact point ($n = 20, 40, 60, 80$). Q-factor $Q_0^{-1} = 0.1$ for a block half-space (Problem 1) and for a layer lying on the half-space (Problem 2). Q-factor $Q_0^{-1} = 0.05$ for the block half-space lying under the layer (Problem 2). In addition, the wave propagation velocity in the block half-space and in the block layer lying on the half-space is two times less than the wave propagation velocity in the half-space lying under the layer.

It can be seen that the reflection and refraction of waves at the lower boundary of the layer leads to a rearrangement of the wave propagating along the surface of the half-space. In the problem for a halfspace, the maximum amplitude of the block velocities is less than in the problem for a layer on a halfspace.

## 7. CONCLUSIONS

Mathematical modeling of the process of wave propagation in a block medium is carried out taking into account internal friction under non-stationary action. The proposed model is new and is based on the idea that the dynamic behavior of a block medium can be roughly described as the movement of rigid blocks due to the compressibility of the interlayers between them, and that the deformation of the interlayers can be approximately described by the Maxwell and Voigt elements.

The use of a rheological model of interlayers between blocks with two Maxwell elements and one Voigt element makes it possible to select the viscosity and stiffness parameters of these elements so that the quality factor of the material deviates from the given constant value by no more than 5% in the frequency range from 4% to 100% of the maximum frequency that is of interest.

Using this model, the Lamb problem is numerically solved, namely, the propagation of seismic waves in a block medium is studied under a non-stationary concentrated vertical action on the surface of a halfspace and on the surface of a layer lying on a half-space. For the Lamb problem, the degree of attenuation of the maximum amplitudes of the block displacement velocities on the surface of a block half-space is studied depending on the quality factor of the material and the distance from the impact point. It is shown that in the problem for a half-space the maximum amplitude of the block velocities is less than in the problem for a layer on a half-space.

In contrast to [8], where the Voigt model was used to simulate the deformation of interlayers, the model proposed in this article with an almost constant quality factor demonstrates a more uniform decay of the perturbation amplitudes for the entire spectrum of frequencies of interest.

The results obtained above show that the presence of a block structure in a medium leads to the following changes in its behavior compared to what is predicted by the model of a homogeneous elastic medium, the mechanical properties of which are obtained by averaging the mechanical properties of a block medium: (a) in contrast to a homogeneous elastic medium, waves propagate with dispersion in a block medium; (b) the propagation velocities of low-frequency longitudinal and Rayleigh waves on the surface of a block medium are much lower than the corresponding velocities in a homogeneous elastic medium; (c) the main contribution to the wave process on the surface of a block medium is made by low-frequency waves in the vicinity of the Rayleigh wave front; (d) behind the front of the Rayleigh wave in a block medium, high-frequency oscillations are observed, which are absent in a homogeneous elastic medium; (e) the propagation velocity of low-frequency waves and the degree of their attenuation are determined by the mass of the blocks, their sizes, and the properties of the interlayers; (f) the dissipative properties of the interlayers lead to additional attenuation on the half-space surface of low-frequency perturbations near the Rayleigh wave front and high-frequency oscillations behind the Rayleigh wave front.

## FUNDING

The work was carried out within the framework of the research project of the Ministry of Science and Higher Education of the Russian Federation (registration number 121062200075-4).

## REFERENCES

1. M. A. Sadovsky, "Natural block size of rock and crustal units," Dokl. Earth Sci. Sect. **247** (4), 22–24 (1979).
2. M. V. Kurlenya, V. N. Oparin, and V. I. Vostrikov, "Pendulum-type waves. Part II: Experimental methods and main results of physical modeling," J. Min. Sci. **32** (4), 245–273 (1996). https://doi.org/10.1007/BF02046215
3. M. V. Kurlenya, V. N. Oparin, and V. I. Vostrikov, et al., "Pendulum waves. Part III: Data of on-site observations," J. Min. Sci. **32** (5), 341–361 (1996).
   https://doi.org/10.1007/BF02046155
4. E. N. Sher and A. G. Chernikov, "Estimate of block medium structure parameters: a model case-study of seismic sounding of a brick wall," J. Min. Sci. **56** (4), 512–517 (2020). https://doi.org/10.1134/S1062739120046800
5. N. I. Aleksandrova, "Elastic wave propagation in block medium under impulse loading," J. Min. Sci. **39** (6), 556–564 (2003).
   https://doi.org/10.1023/B:JOMI.0000036223.58270.42
6. M. V. Kurlenya, V. N. Oparin, and V. I. Vostrikov, "Formation of elastic wave packets as a result of pulsed excitation of block media. Pendulum-type waves V$\mu$," Dokl. Akad. Nauk SSSR, **333** (4), 3–13 (1993).

7. N. I. Aleksandrova, A. G. Chernikov, and E. N. Sher, "Experimental investigation into the one-dimensional calculated model of wave propagation in block medium," J. Min. Sci. **41** (3), 232–239 (2005).
   https://doi.org/10.1007/s10913-005-0088-y
8. N. I. Aleksandrova, "Seismic waves in a three-dimensional block medium," Proc. R. Soc. A. **472** (2192), 20160111 (2016).
   https://doi.org/10.1098/rspa.2016.0111
9. N. I. Aleksandrova, "Pendulum waves on the surface of block rock mass under dynamic impact," J. Min. Sci. **53** (1), 59–64 (2017).
   https://doi.org/10.1134/S1062739117011847
10. V. M. Sadovskii and O. V. Sadovskaya, "Supercomputer modeling of wave propagation in blocky media accounting fractures of interlayers," in *Nonlinear Wave Dynamics of Materials and Structures, Advanced Structured Materials,* Vol. 122, Ed. by H. Altenbach, V. Eremeyev, I. Pavlov, and A. Porubov (Springer, 2020), pp. 379–398.
    https://doi.org/10.1007/978-3-030-38708-2_22
11. V. M. Sadovskii and O. V. Sadovskaya, "Numerical algorithm based on implicit finite-difference schemes for analysis of dynamic processes in blocky media," Russ. J. Num. Anal. Math. Modell. **33** (2), 111–121 (2018).
    https://doi.org/10.1515/rnam-2018-0010
12. C. M. Zener, *Elasticity and Anelasticity of Metals*, 1st. ed. (Univ. of Chicago, Chicago, 1948).
13. M. A. Biot, "Theory of stress-strain relations in anisotropic viscoelasticity and relaxation phenomena," J. Appl. Phys. **25**, 1385–1391 (1954).
    https://doi.org/10.1063/1.1721573
14. Y. C. Fung, *Fundations of Solid Mechanics* (Prentice Hall, 1965).
15. J. Bielak, H. Karaoglu, and R. Taborda, "Memory-efficient displacement-based internal friction for wave propagation simulation," Geophys. **76** (6), T131–T145 (2011).
    https://doi.org/10.1190/geo2011-0019.1
16. M. Toksöz and D. Johnston, *Seismic Wave Attenuation* (Soc. of Exploration Geophysicists, Tulsa, Okla., 1981).
17. E. Kjartansson, "Constant Q-wave propagation and attenuation," J. Geophys. Res. **84**, 4737–4748 (1979).
    https://doi.org/10.1029/JB084iB09p04737